# Resource Aware Min-Min (RAMM) Algorithm for Resource Allocation in Cloud Computing Environment

Syed Arshad Ali, Student Member, IEEE, Mansaf Alam, Member, IEEE
Department of Computer Science
Jamia Millia Islamia
New Delhi, India
arshad158931@st.jmi.ac.in, malam2@jmi.ac.in

*Abstract* — Resource allocation (RA) is a significant aspect in Cloud Computing, which facilitates the Cloud resources to Cloud consumers as a metered service. The Cloud resource manager is responsible to assign available resources to the tasks for execution in an effective way that improves system performance, reduce response time, reduce makespan and utilize resources efficiently. To fulfil these objectives, an effective Tasks Scheduling algorithm is required. The standard Min-Min and Max-Min Task Scheduling Algorithms are available, but these algorithms are not able to produce better makespan and effective resource utilization. This paper proposed a Resource Aware Min-Min (RAMM) Algorithm based on classic Min-Min Algorithm. The RAMM Algorithm selects shortest execution time task and assign it to the resource which takes shortest completion time. If minimum completion time resource is busy then the RAMM Algorithm selects next minimum completion time resource to reduce waiting time of task and better resource utilization. The experiment results show that the RAMM Algorithm produces better makespan and load balance than standard Min-Min, Max-Min and improved Max-Min Algorithms.

## I. Introduction

Cloud Computing provides Internet based dynamic computing services using large-scale virtualized resources. Cloud Computing is a combination of parallel and distributed computing [1]. It serves distributed computing resources to globally located users simultaneously to deliver resource scalability, economic use of resources and on-demand services [2]. Resource allocation in Cloud Computing is very complex because of dynamic nature of Cloud environment [3]. The user's demand changes dynamically and availability of resources are also changing very frequently. From the Cloud provider's point of view, a huge amount of resources is needed to allocate among the globally distributed Cloud users dynamically in a cost-effective way while from the Cloud user's perspective a reliable and economic computing services are needed on demand [4]. There must be a Service Level Agreement (SLA) between the Cloud provider and the Cloud user which consider multiple parameters

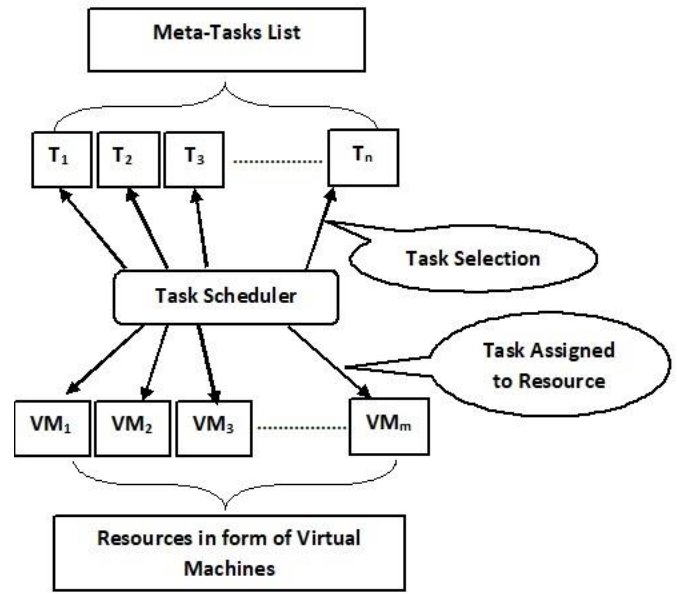

Fig 1: Tasks-Resource Mapping by Cloud-Scheduler

like the cost of service, the completion time of service (makespan), and throughput etc. [5].

In Cloud Computing, resource allocation is the process of allocating virtual machines, (storage, computing, and networking) to the Cloud user's applications. Cloud resource allocation comprises both Cloud resource provisioning and scheduling. Cloud Computing mainly relies on virtualization, which enables a physical device to be virtually distributed into one or more virtual machines (VMs) [6]. Virtual machines are used for computation of user applications. Due to virtualization, unutilized resources of physical machines can be further used by another virtual machine to speed up the tasks execution and resource utilization [7]. Resource allocation strategy should overcome the problems related to over and under provisioning of resources resource, scarcity of resources, contention, and fragmentation of resources [8].

Scheduling is an important aspect of any computing system. In general, CPU scheduling deals with the execution of user-submitted jobs. First, all the user submitted jobs wait in ready queue for their turn of execution. The time spends in the ready queue by the job is known as waiting time. CPU scheduler selects jobs from the ready queue based on some criteria fulfill by the job and assign to the CPU for execution [9]. The waiting time depends on several factors including resource availability, the priority of the job, the load on the system. Total time for execution of all jobs is known as makespan. The scheduling process should minimize the makespan to improve the system performance. Cloud user submits a task to Cloud scheduler which is responsible to select the available Virtual Machine and allocate it to the user's submitted task to fulfil Cloud user and Cloud provider's requirements effectively. Fig-1 illustrate how Cloud scheduler schedule user's task to available resources. Various Task Scheduling algorithms consider task completion time and task execution time as scheduling criteria for resource allocation to user's tasks. The Max-Min [10], Min-Min [11], RASA [12] and improved Max-Min [13] algorithms also use these scheduling criteria for resource allocation. In Min-Min algorithm the smallest completion time task schedule first to the fastest execution time resource. The major drawbacks of the Min-Min algorithm are load imbalanced and starvation of tasks with large service time. To solve these problems, Max-Min algorithm is proposed which schedule largest completion time task to smallest execution time resource. When the numbers of small tasks are more than large tasks, then Max-Min seems a better choice for Task Scheduling. But in some cases, if large tasks are more than small tasks then total completion time and throughput of the system decreases. One more resource aware scheduling algorithm (RASA) has been proposed for task scheduling which combines the features of both Max-Min and Min-Min algorithms. In this algorithm, author also used completion time for each task and apply Max-Min and Min-Min algorithms one after another according to the number of resources available if number of resources are even then it applies Max-Min else it applies Min-Min algorithm. Another improved Max-Min algorithm proposed a different method than these algorithms, it schedules large execution time task first to the minimum completion time resource compared to basic Max-Min which assign large completion time task first to the minimum execution time resource.

In this paper, we proposed another Task Scheduling algorithm, Resource Aware Min-Min (RAMM) algorithm based on Min-Min algorithm. We use the concept of Min-Min algorithm in such a way, that schedule minimum execution time task to a resource with minimum completion time instead of basic Min-Min algorithm which select smallest completion time task to the smallest execution time resource. The basic Min-Min algorithm suffers from the load imbalance problem because it always selects the resource which has minimum execution time for the task, if that resource is not ready then the task must wait for it while rest of resources are idle in this situation, which causes resource's load imbalance and increased makespan in Cloud system. While in this proposed RAMM algorithm if minimum completion time resource is busy then it will select next minimum completion time resource for that task which makes load balance among resources and decrease makespan because no task will wait for busy resource if next minimum completion time resource is available.

Rest of the paper comprises following sections. In Section-II related work is discussed, in section-III proposed algorithm is described. Implementation of algorithm has been done in section-IV while experimental setup is described in section-V. Section-VI and VII illustrate some experiment examples and results respectively. Section-VIII has conclusion and future work.

## II. RELATED WORK

Resource allocation is a NP-hard problem in Cloud computing. Task scheduling is one of the important aspect of resource allocation. Task scheduling refers the allocation of tasks to the available resources in an efficient way that fulfil task requirement and utilize resources efficiently. Many researchers are working in this field and proposed many Task Scheduling algorithms which can fulfill the need for Cloud environment.

In [14] author uses Min-Min algorithm to propose a user's priority based Min-Min scheduling algorithm. The focus of this algorithm is on user's priority to fulfill the Service Level Agreement (SLA). It prioritized user task to overcome the unbalanced workload problem of the basic Min-Min algorithm.

In Max-Min algorithm small jobs starved due to the priority given to large jobs, to overcome this problem in [15], the author proposed an algorithm named as Max-Min spare time (MMST), which reduces waiting time of small jobs and utilizes resource efficiently. The algorithm also reduces service cost of Cloud resources.

In [16] author used the improved Max-Min algorithm as a base to propose an enhanced Max-Min algorithm with some changes to improve the overall makespan and load imbalanced problem. Instead of the scheduling largest execution time task to resource produces minimum completion time, it assigns average execution time task to minimum completion time resource.

A new enhanced load balancing algorithm is developed in [17] based on load balanced Min-Min algorithm. Load balanced Min-Min algorithm works in two steps, in the first step it applies Min-Min algorithm and in the second step it rescheduled tasks to unutilized resources to improve makespan as well as resource load balancing but sometimes it does not give appropriate results because it schedules task with minimum completion time. On the other hand, enhanced load balanced Min-Min algorithm also works in two steps, Min-Min algorithm is applied in first step and in the second step it reschedules the largest completion time task to suitable resource for better resource utilization and to improve makespan.

Cloud Computing is more popular due to its elastic property, a user can expand or reduce his infrastructure resources according to his requirement using Cloud Computing. In [18], an improved Max-Min algorithm for elastic Cloud is proposed which balance the work load among the resources by maintaining two tables. The task-status table maintains the expected completion time of each task and virtual machine status table maintains the estimated load of each virtual machine. The algorithm works in two phases, in first phase it executes VM task estimation algorithm and in second phase it executes task allocation algorithm. It improves the resource utilization and response time of tasks.

In [13], an improved Max-Min algorithm is proposed based on a basic Max-Min algorithm. The Max-Min algorithm gives priority to large task for execution by assigning them to fastest resource and small tasks are executed concurrently by other resources. If small tasks are more than large tasks, then concurrent execution is not possible which increase makespan. To overcome this problem an improved Max-Min algorithm is proposed which schedule maximum execution time task to minimum completion time resource. NBDMMM Algorithm [23] have improvises allocation of resources in a virtualized cloud. The work [24] provides analysis of resource usage and an attempt to give an insight about benevolent of production trace related to the ones in cloud environment

## III. RESOURCE AWARE MIN-MIN (RAMM) ALGORITHM

The proposed algorithm is an advance version of traditional Min-Min Task Scheduling algorithm for grid system. Smallest completion task scheduled first to the fastest execution time resource in basic Min-

Min algorithm. If fastest (minimum execution time) resource is busy in executing another task then that minimum completion time task will have to wait until that resource will ready for it, this means waiting time of the task will increase which also increase makespan and decrease the system performance. The proposed resource aware Min-Min (RAMM) algorithm works differently, it schedules the minimum execution time task to the minimum completion time resource. In addition, it also assigns the next minimum completion time resource to that task, if the minimum completion time resource is busy in executing another task, which reduces the waiting time of the task and increase the makespan. Another problem with basic Min-Min algorithm is that if all tasks have minimum execution time on a single resource then all tasks will assign to that resource which increase the load of that resource while rest of the resources are idle at that time, which causes load imbalance problem and decrease efficient resource utilization. The proposed algorithm overcomes the basic Min-Min load imbalance problem by effective use of the idle or available resources. Several Scheduling criteria can be considered for performance enhancement in Cloud Computing systems [19]. Cloud's system performance can be measured in resource utilization, throughput, load-balancing, system usage, turnaround time, response time, waiting time and some other criteria could also be considered for characterizing the Cloud system's performance such as user priority, quality of service (QoS), resource failure etc. One of the most prevalent and extensively considered Scheduling criteria in Cloud resource allocation is the minimization of makespan [20]. Makespan is the time taken by the system to complete all submitted tasks. Smaller makespan value shows that the scheduler is providing efficient scheduling of the tasks to the system resources. In this proposed algorithm author considers the following aspects of the tasks for Task Scheduling. The flow-chart of the proposed algorithm is given in fig-2. The main concern of this proposed algorithm is to reduce the makespan and balance the load of resources for improving the performance of the system.

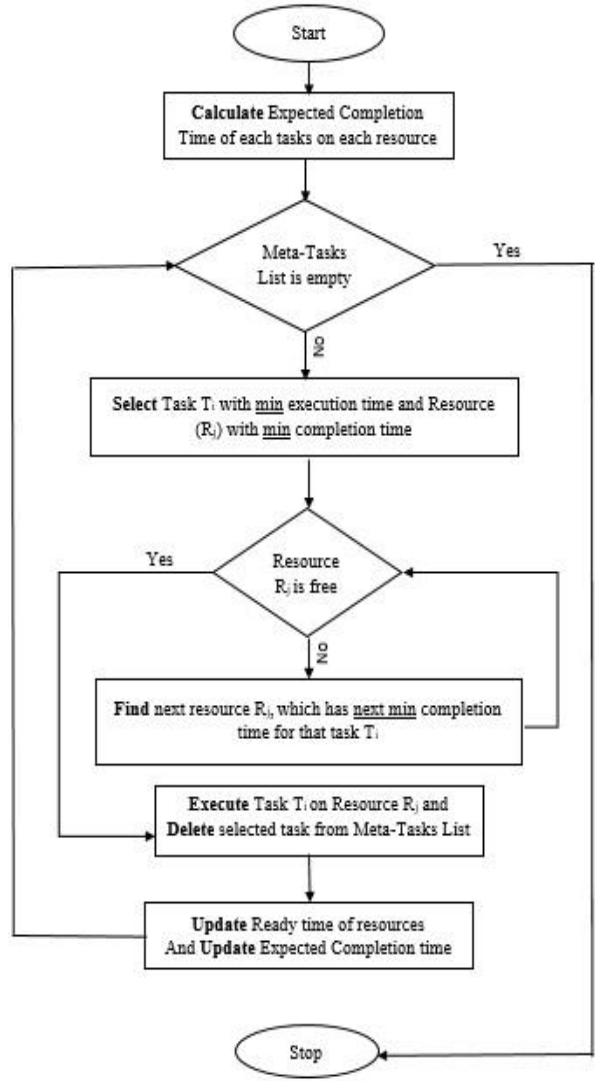

Fig 2: Flowchart of Resource Aware Min-Min (RAMM) algorithm

**Expected execution time:** $ET_{ij}$ is a unit of time taken by the resource $R_j$ to complete task $T_i$ when resource $R_j$ has no previous task for execution. It is also known as burst time of task $T_i$.

**Expected completion time:** $CT_{ij}$ is the measure of time taken by resource $R_j$ to complete task $T_i$ including ready time of resource $R_j$ allocated to the previously assigned task. Expected completion time is the sum of expected execution time and ready time i.e. $CT_{ij} = ET_{ij} + RT_j$.

## IV. IMPLEMENTATION OF ALGORITHM

Let m be the number of resources which must process n number of tasks. The scheduler is responsible for allocating each task to one or more resources for their execution. We have developed an algorithm for improving the performance of the system by decreasing makespan of the tasks. In the proposed algorithm we have considered T as the set of all tasks, T = {$T_1$, $T_2$, $T_3$, $T_4$, ……… $T_n$} and R as the set of all resources which will be mapped to tasks, R = {$R_1$, $R_2$, $R_3$, $R_4$, ……...$R_m$}. The Cloud scheduler has the expected execution time of each task on each resource, which is given in form of matrix $ET_{ij}$. The algorithm calculates the expected completion time ($CT_{ij}$) of each tasks $T_i$ on each resource $R_j$, each resource may have some previously assigned task for execution which takes some time to get ready this resource for coming task $T_i$, this amount of time is known as ready time of resource $R_j$ or waiting time of task $T_i$ which is denoted as RT = {$RT_1$, $RT_2$, $RT_3$ …………$RT_j$}, therefore each task's expected execution time will be added to the waiting time of each resource to get the expected completion time on each resource. Initially all the resources are free that's why initial expected task completion time is same as expected task execution time. If small tasks are more than large tasks in Min-Min algorithm, then the fast computing resources will always busy and slow computing resources are idle most of the time which causes starvation for large tasks and arises problem of load imbalance because it does not support concurrent execution of the tasks. Therefore, the traditional Task Scheduling algorithms are not well suited in Cloud environment due to its dynamic nature. To overcome these problems of starvation and unbalance load, we use an alternate method in which no task will have to wait if there is any available resource which support simultaneous execution of the tasks. When task will complete its execution, it will remove from the meta-task sets. And the expected completion time of remaining task on each resource will be updated for further scheduling and this process will repeat until all tasks complete their execution. The pseudo code of the proposed algorithm is represented in Fig-3. The expected completion time matrix is denoted by $CT_{ij}$ which is sum of expected execution time ($ET_{ij}$) and ready time ($RT_j$) of resource $R_j$.

---

**Input:** Meta-Tasks List and Set of Resources.

**Output:** Mapping of Tasks to the system Resources for execution.

1. for all tasks $T_i$ in Meta-Tasks List
2.    for all resources $R_j$
3.       Calculate $CT_{ij} = ET_{ij} + RT_j$
4.    end
5. end
6. do until all tasks in Meta-Tasks List are mapped
7.    find the task $T_k$ with the minimum Execution time and the resource $R_l$ that gives minimum completion time
8.    if Resource $R_l$ is busy then
9.       find the next resource $R_l$ with the next minimum completion time
10.       goto step-8
11.    else
12.       Execute task $T_k$ on the resource $R_l$
13.    end if
15.    delete task $T_k$ from Meta-Tasks List
16.    update $RT_l$
17.    update $CT_{il}$ for all i
18. end do

Fig 3: Resource Aware Min-Min (RAMM) algorithm

## V. EXPERIMENTAL SETUP

The experiment work for the proposed algorithm is done using CloudSim [21], a simulator to simulate and model Cloud Computing system and application environment. CloudSim provide both system and working modelling of Cloud infrastructures such as Cloud data centers, Cloud resources (VMs), cloudlets and resource provisioning and scheduling policies. Different problem samples are used to illustrate the results of proposed algorithm. The Intel Core i5 system with 12 GB of RAM is used for experimental work. Some graphical representation is done using MATLAB parallel computing toolbox [22].

## VI. EXPERIMENT ON PROPOSED ALGORITHM

To get a practical experiment on our proposed algorithm, we assume there are four tasks {T = T1, T2, T3 and T4} and two resources {R = R1 and R2} in the system. Table-1 has Meta-Tasks requirement contains instruction volume and data volume of each task which must be scheduled for execution. Table-2 represents resource specification including processing speed and bandwidth of the resource, here resources are basically virtual machines which must be scheduled by the proposed algorithm to the user submitted task for their execution.

TABLE I   META-TASKS REQUIREMENT (MTR)

| Task | Instruction Volume (MI) | Data Volume (Mb) |
|---|---|---|
| Task(T1) | 256 | 88 |
| Task(T2) | 35 | 31 |
| Task(T3) | 327 | 96 |
| Task(T4) | 210 | 590 |

TABLE 2   RESOURCE SPECIFICATIONS (RS)

| Resource | Processing Speed (Mips) | Bandwidth (Mbps) |
|---|---|---|
| R1 | 150 | 300 |
| R2 | 300 | 15 |

Table-3 shows the expected execution time of each task (Ti) on each resource (Rj). This is calculated by the given data of Table-1 and Table-2. Expected execution time is calculated by the given formula.

$$ET_{ij} = [(MI_i \div MIPS_j) + (Mb_i \div Mbps_j)]$$

Initially, the expected completion time (CT) of each task is equal to the expected execution time (ET), because initially all the resources are in idle state i.e. ready time (RT=0) for all the resources. After first iteration, the given formula is used to calculate the expected completion time of each task $T_i$ on each resource $R_j$.

$$CT_{ij} = ET_{ij} + RT_j$$

In the first iteration the proposed algorithm selects task $T_2$ which has minimum execution time for execution on the resource $R_1$ which has smallest completion time, at the same time resource $R_2$ is free and it has the next minimum completion time for the task $T_1$ so task $T_1$ will schedule to the resource $R_2$ to make concurrent execution of the tasks and improve load balance among the resources.

TABLE 3   COMPLETION TIME

| Task/Resource | R1 | R2 |
|---|---|---|
| Task(T1) | 2 | 6 |
| Task(T2) | 1 | 3 |
| Task(T3) | 3 | 8 |
| Task(T4) | 3 | 40 |

Now both resources are busy in executing the tasks. The resource $R_1$ will get free after 1 unit of time and at that time resource $R_2$ is remain busy for executing task $T_1$. Now we calculate expected completion time of remaining tasks and assign minimum execution time task $T_3$ to the resource $R_1$. In next iteration the resource $R_1$ will free and resource $R_2$ is still busy in execution of task $T_1$, therefore task $T_4$ will assign to the resource $R_1$ and complete its execution. The Gantt-Chart is shown in Fig-4, which represents the execution order of the tasks. From the figure we can see that the makespan of the system using proposed Resource aware Min-Min (RAMM) algorithm is 7 units. And both resources share the load of system equally.

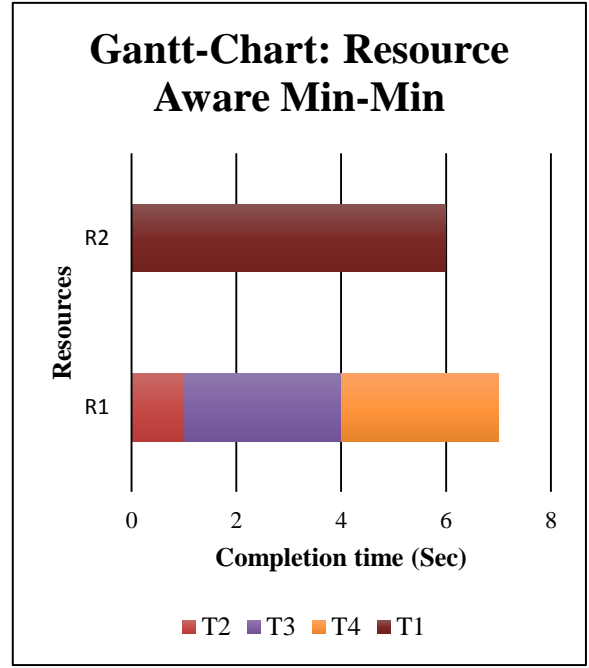

Fig 4: Gantt-Chart of Resource Aware Min-Min Algorithm

## VII. RESULT AND DISCUSSION

In the previous section, we have find out the makespan of the tasks as 7 unit of time using proposed RAMM algorithm. Now in this section we will evaluate the makespan for Min-Min, Max-Min and improved Max-Min algorithms and compared with the proposed algorithm. We have performed these algorithms one by one on the given data in Table-3. In Fig-5 execution of Min-Min algorithm is shown. The makespan of Min-Min algorithms is 9 unit of time and only one resource $R_1$ is busy while resource $R_2$ is idle all the time which makes load imbalance. Fig-6 shows the Gantt-Chart of Max-Min algorithms which executes larger task first and the makespan using Max-Min algorithm is same as Min-Min algorithm i.e. 9 unit and it also suffers from the load imbalance problem. Now we applied improved Max-Min algorithm which gives makespan as 8 unit and load is balanced Fig-7 shows the Gantt-Chart of improved Max-Min algorithm. Now from these results we can see that proposed RAMM algorithm is far better than Min-Min and Max-Min algorithm in both aspects makespan and load balance. Proposed RAMM algorithm and improved Max-Min algorithm balance the load among the resources and execute tasks concurrently but proposed RAMM algorithm gives less makespan than improved Max-Min therefore we can say that our proposed RAMM algorithm is better than all these algorithms and gives better result.

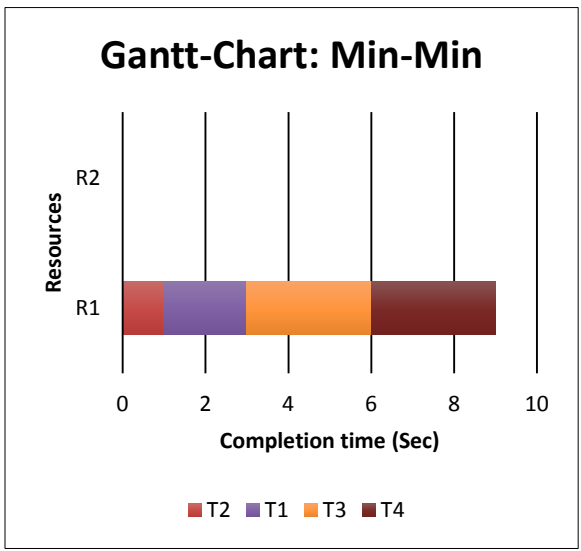

Fig 5: Gantt-Chart of Min-Min Algorithm

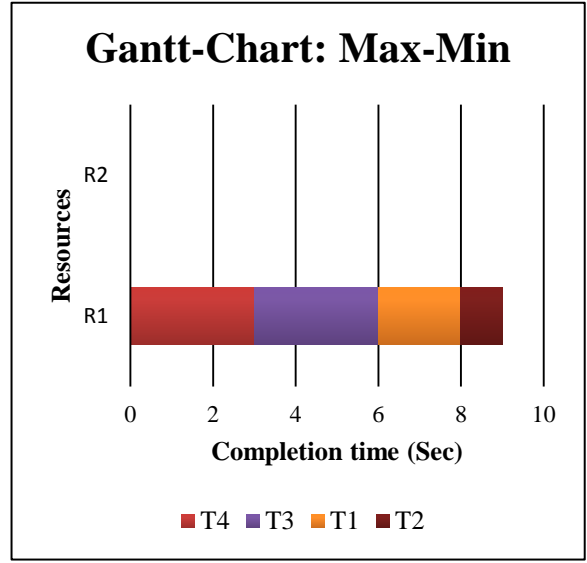

Fig 6: Gantt-Chart of Max-Min Algorithm

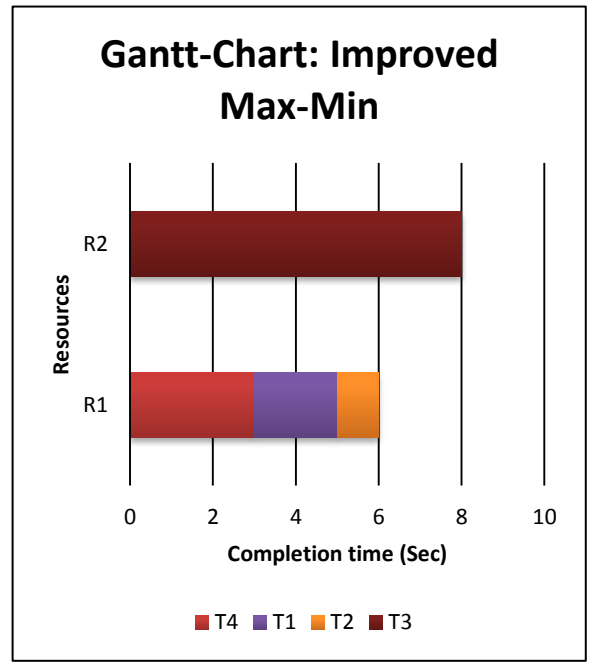

Fig 7: Gantt-Chart of Improved Max-Min Algorithm

To validate these results, we are taking three more set of problem samples. Table-4 has Meta-Tasks requirement of three problem samples and Table-5 has resource specification of these problem sample. Table-6 consists the initial completion time tasks. Min-Min, Max-Min, improved Max-Min and proposed Resource Aware Min-Min algorithms are executed on these problem sample. The makespan has been calculated and shown in Table-7. The fig 8 shows the comparison of makespan of these algorithms. The proposed algorithm (RAMM) gives better makespan in all three-problem sample than other algorithms.

TABLE 4    META-TASKS REQUIREMENT (MTR)

| Problem | Task | Instruction Volume (MI) | Data Volume (Mb) |
|---|---|---|---|
| P1 | Task(T1) | 256 | 88 |
|  | Task(T2) | 35 | 31 |
|  | Task(T3) | 327 | 96 |
|  | Task(T4) | 210 | 590 |
| P2 | Task(T1) | 128 | 44 |
|  | Task(T2) | 69 | 62 |
|  | Task(T3) | 218 | 94 |
|  | Task(T4) | 21 | 59 |
| P3 | Task(T1) | 88 | 20 |
|  | Task(T2) | 31 | 350 |
|  | Task(T3) | 100 | 207 |
|  | Task(T4) | 50 | 21 |

TABLE 5    RESOURCE SPECIFICATIONS (RS)

| Problem | Resource | Processing Speed (Mips) | Bandwidth (Mbps) |
|---|---|---|---|
| P1 | R1 | 150 | 300 |
|  | R2 | 300 | 15 |
| P2 | R1 | 50 | 100 |
|  | R2 | 100 | 5 |
| P3 | R1 | 300 | 300 |
|  | R2 | 30 | 15 |

TABLE 6    COMPLETION TIME

| Problem | Task | R1 | R2 |
|---|---|---|---|
| P1 | Task(T1) | 2 | 6 |
|  | Task(T2) | 1 | 2 |
|  | Task(T3) | 3 | 8 |
|  | Task(T4) | 3 | 40 |
| P2 | Task(T1) | 3 | 10 |
|  | Task(T2) | 2 | 13 |
|  | Task(T3) | 5 | 21 |
|  | Task(T4) | 1 | 12 |
| P3 | Task(T1) | 1 | 7 |
|  | Task(T2) | 1 | 14 |
|  | Task(T3) | 1 | 14 |
|  | Task(T4) | 1 | 4 |

TABLE 7    COMPARISON OF MAKESPAN OF ALGORITHMS

| Problem | Min-Min Algorithm | Max-Min Algorithm | Imp Max-Min Algorithm | Resource Aware Min-Min Algorithm |
|---|---|---|---|---|
| P1 | 9 | 9 | 8 | 7 |
| P2 | 11 | 11 | 13 | 10 |
| P3 | 4 | 4 | 14 | 4 |

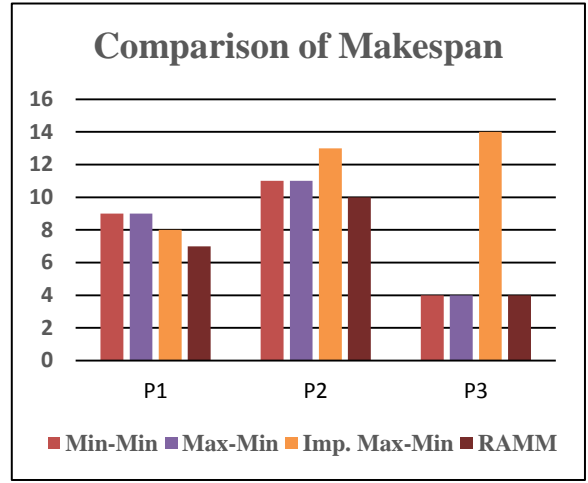

Fig 8: Makespan Comparison of different scheduling Algorithm

VIII.    CONCLUSION AND FUTURE WORK

Resource allocation in Cloud Computing is an important aspect. Due to the novelty of Cloud Computing many Task Scheduling algorithms have been proposed in this paper we proposed a Resource Aware Min-Min (RAMM) algorithm for Cloud environment. Various previous proposed algorithms like Min-Min, Max-Min and improved Max-Min have been compared with the RAMM algorithm. The proposed algorithm gives better results in the form of Minimum makespan and effective resource load balancing. The resources were not utilized in the Min-Min and Max-Min algorithm which causes increase in waiting time of tasks and results in high makespan of the system. While in proposed resource aware Min-Min algorithm the main concern is to reduce makespan and resource load balancing as the results shows the proposed algorithm gives better makespan and resource load balancing.

The proposed algorithm works for single cloud environment in which resources are present in a single Cloud. In the future this algorithm will expand for the multi-Cloud environment in which resources (Virtual Machines) are in multiple Clouds environment.

ACKNOWLEDGEMENT

This work was supported by a grant from "Young Faculty Research Fellowship" under Visvesvaraya PhD Scheme for Electronics and IT, Department of Electronics & Information Technology (DeitY), Ministry of Communications & IT, Government of India.